\documentclass[11pt]{article}

\usepackage{amstext,amsgen,latexsym,amsmath}
\usepackage{amstext,amssymb,amsfonts,latexsym}
\usepackage{theorem} \usepackage{pifont}
\usepackage[dvips]{graphics,epsfig}

 \newcommand{\bs}{\bigskip} \newcommand{\ms}{\medskip}
 \newcommand{\n}{\noindent} 
 \newcommand{\hs}[1]{\hspace*{ #1 mm}}

\theoremstyle{plain}
\theoremheaderfont{\bfseries}
 \newtheorem{theorem}{Theorem}[section]
 \newtheorem{lemma}[theorem]{Lemma}

 \newenvironment{proof}{\par \noindent
            {\bf Proof. \hs{2}}}{\hfill$\Box$ \vspace*{3mm}}

 \newcommand{\bra}[1]{\langle #1 |}
 \newcommand{\ket}[1]{| #1 \rangle}

\newcommand{\ignore}[1]{}

\begin{document}
\pagestyle{plain}

\begin{center}
{\Large {\bf $(4,1)$-Quantum Random Access Coding Does Not Exist}}
\bs\\

\begin{center}
{\sc Masahito Hayashi}$^1$ \hspace{10mm} {\sc Kazuo Iwama}$^2$ \hspace{10mm} {\sc Harumichi Nishimura}$^3$\\
{\sc Rudy Raymond}$^2$ \hspace{30mm} {\sc Shigeru Yamashita}$^4$ 
\end{center}

\

$^1${ERATO-SORST Quantum Computation and Information Project,\\ Japan Science and Technology Agency}

{\tt masahito@qci.jst.go.jp}

$^2${Graduate School of Informatics, Kyoto University}

{\tt $\{{\tt iwama, raymond}\}$@kuis.kyoto-u.ac.jp}

$^3${Graduate Scool of Science, Osaka Prefecture University}\footnote{This work was done while HN was 
in Graduate School of Informatics, Kyoto university.}

{\tt hnishimura@mi.s.osakafu-u.ac.jp}

$^4${Graduate School of Information Science, Nara Institute of Science and Technology} 

{\tt ger@is.naist.jp}
\end{center}
\bs

\n{\bf Abstract.}\hs{1} An $(n,1,p)$-Quantum Random Access (QRA) coding, introduced by Ambainis, 
Nayak, Ta-shma and Vazirani in \emph{ACM Symp. on Theory of Computing }1999, is the following communication 
system: The sender which has $n$-bit information encodes his/her 
information into one qubit, which is sent to the receiver. The receiver 
can recover any one bit of the original $n$ bits correctly with 
probability at least $p$, through a certain decoding process based 
on positive operator-valued measures. Actually, Ambainis et al. shows the existence of a 
$(2,1,0.85)$-QRA coding and also proves the impossibility of its classical  
counterpart. Chuang immediately extends it to a $(3,1,0.79)$-QRA coding 
and whether or not a $(4,1,p)$-QRA coding such that $p > 1/2$ exists has 
been open since then. This paper gives a negative answer to this open question. 

\ms

\section{Introduction}
The state of $n$ quantum bits (qubits) is given by a vector of length 
$2^n$ and seems to hold much more information than (classical) $n$ bits. 
However, due to the famous Holevo bound \cite{Hol79}, this is not true 
information-theoretically, i.e., we need $n$ qubits to transmit $n$-bit 
information faithfully. As an interesting challenge to this most basic 
fact in quantum information theory, Ambainis, Nayak, Ta-shma and 
Vazirani introduced the notion of {\em quantum random access (QRA) coding} \cite{ANTV99}. 
(The paper \cite{ANTV02} includes the contents of \cite{ANTV99} and their improvement in \cite{Nay99}.) 
They explored the possibility of using much less qubits if the receiver has to recover only {\it partial} 
bits, say one bit out of the $n$ original ones, which are not known by the sender in advance. 

As a concrete example, they give $(2,1,0.85)$-QRA coding; the sender 
having two-bit information sends one qubit and the receiver can recover 
any one of the two bits with probability at least $0.85$. It is also 
proved that this is not possible classically, i.e., if the sender can 
transmit one classical bit, then the success probability is at most 
$1/2$. This $(2,1,0.85)$-QRA coding is immediately extended to 
$(3,1,0.79)$-QRA coding by Chuang (as mentioned in \cite{ANTV99}) and it 
has been open whether we can make a further extension (i.e., whether 
there is an $(n,1,p)$-QRA coding such that $n \ge 4$ and $p > 1/2$) since then. 

{\bf Our Contribution~} This paper gives a negative answer to this open question, 
namely, we prove there is no $(4,1,p)$-QRA coding such that $p$ is strictly greater than $1/2$. 
Our proof idea is to use the well-known geometric fact that a three-dimensional ball cannot be 
divided into $16$ nonempty regions by four planes. (Interestingly, the 
proof for the non-existence of a classical counterpart of $(2,1,p)$-QRA 
coding in \cite{ANTV99} uses a similar geometric fact, i.e., a straight 
line cannot stab all insides of the four quarters of a two-dimensional 
square.) Our result has nice applications to the analysis of {\it quantum 
network coding} which was introduced very recently \cite{HINRY06}. 

In general, the sender is allowed to send $m$ ($\ge 1$) qubits; 
such a system is denoted by $(n,m,p)$-QRA coding. Our result can be extended 
to this general case, namely we can show that $(2^{2m},m,p)$-QRA coding with $p>1/2$ does not exist. 

{\bf Related Work~} For the relation among these three parameters of $(n,m,p)$-QRA coding, 
the following bound is known \cite{Nay99}: $m\geq (1-H(p))n$, where $H$ is the binary entropy function, 
and it is also known \cite{ANTV99} that $(n,m,p)$-QRA coding with $m=(1-H(p))n+O(\log{n})$ exists (which is 
actually classical).  Thus, this bound is asymptotically tight and has 
many applications such as proving the limit of quantum finite automata 
\cite{ANTV99,Nay99}, analyzing quantum communication complexity \cite{BW01,Kla00}, 
designing locally decodable code \cite{KW03,WW05}, and so on. However, it says 
almost nothing for small $n$ and $m$; if we set $n = 4$ and $p > 1/2$, 
for example, the bound implies only $m > 0$. 
This bound neither implies the limit of $n$ for a given $m$ 
if $\epsilon=p-1/2$ is very small, say $\epsilon=1/g(n)$ for rapidly increasing $g(n)$. 
Our second result says that there does exist a limit of $n$ for any small $\epsilon$. 

K\"{o}nig, Maurer and Renner \cite{KMR05} extended the concept of QRA 
coding to the situation that the receiver wants to compute a (randomly 
selected) function on the bits the sender has, and applied their limit 
of its extended concept to the security of the privacy amplification, a  
primitive of quantum key distribution. The study on QRA coding for more 
than two parties was done by Aaronson \cite{Aar05}, who explored the QRA 
coding in the setting of the Merlin-Arthur games.  

\section{Quantum Random Access Coding}
The following definition is due to \cite{ANTV99}. 

{\bf Definition.} An {\em $(n,1,p)$-QRA coding} is a function that maps $n$-bit strings $x \in \{0,1\}^n$ 
to $1$-qubit states $\pmb{\rho}_x$ satisfying the following: 
For every $i\in\{1,2,\ldots,n\}$ there exists a positive operator--valued measure (POVM) 
$E^{i} = \{E_0^{i},E_1^{i}\}$ such that 
$\mathrm{Tr}(E_{x_i}^{i}\pmb{\rho}_x)\geq p$ for all $x\in\{0,1\}^n$, 
where $x_i$ is the $i$-th bit of $x$. 

Recall that a POVM $\{E_0^i, E_1^i\}$ has to satisfy the following 
conditions: ($i$) $E_0^i$ and $E_1^i$ are both nonnegative Hermitian and 
($ii$) $E_0^i + E_1^i = I$. It is well-known, since $E_0^i$ and $E_1^i$ 
are of rank at most $2$, that $E_0^i$ can be written as $E_0^i = \alpha_1 
\ket{u_1}\bra{u_1} + \alpha_2 \ket{u_2}\bra{u_2}$ for some orthonormal basis $\{\ket{u_1},\ket{u_2}\}$, 
$0 \le \alpha_1 \le 1$ and $0 \le \alpha_2 \le 1$. Hence, by ($ii$), $E_1^i = I - E_0^i = 
(1-\alpha_1)\ket{u_1}\bra{u_1} + (1-\alpha_2)\ket{u_2}\bra{u_2}$. If 
$E_0^i$ and $E_1^i$ can be written as $E_0^i = \ket{u_1}\bra{u_1}$ and 
$E_1^i = \ket{u_2}\bra{u_2}$, then the measurement by the POVM $\{E_0^i,E_1^i\}$ is called a 
{\it projective measurement (in the basis $\{\ket{u_1},\ket{u_2} \}$)}.   

We next review $(2,1,0.85)$- and $(3,1,0.79)$-QRA codings. 
\begin{figure}[tbh]
\begin{center}
\includegraphics*[width=6cm]{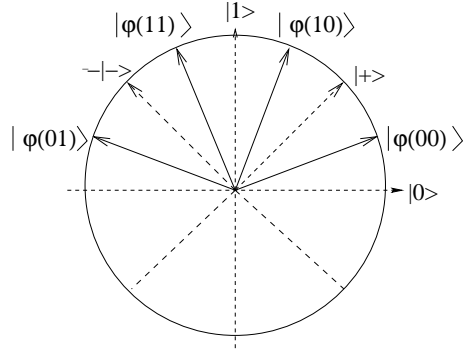}
\caption{The $(2,1,0.85)$-QRA coding in $\ket{0}$-$\ket{1}$ plane 
representation}\label{twotoone}
\end{center}
\end{figure}

\begin{figure}[tbh]
\begin{center}
\includegraphics*[width=7cm]{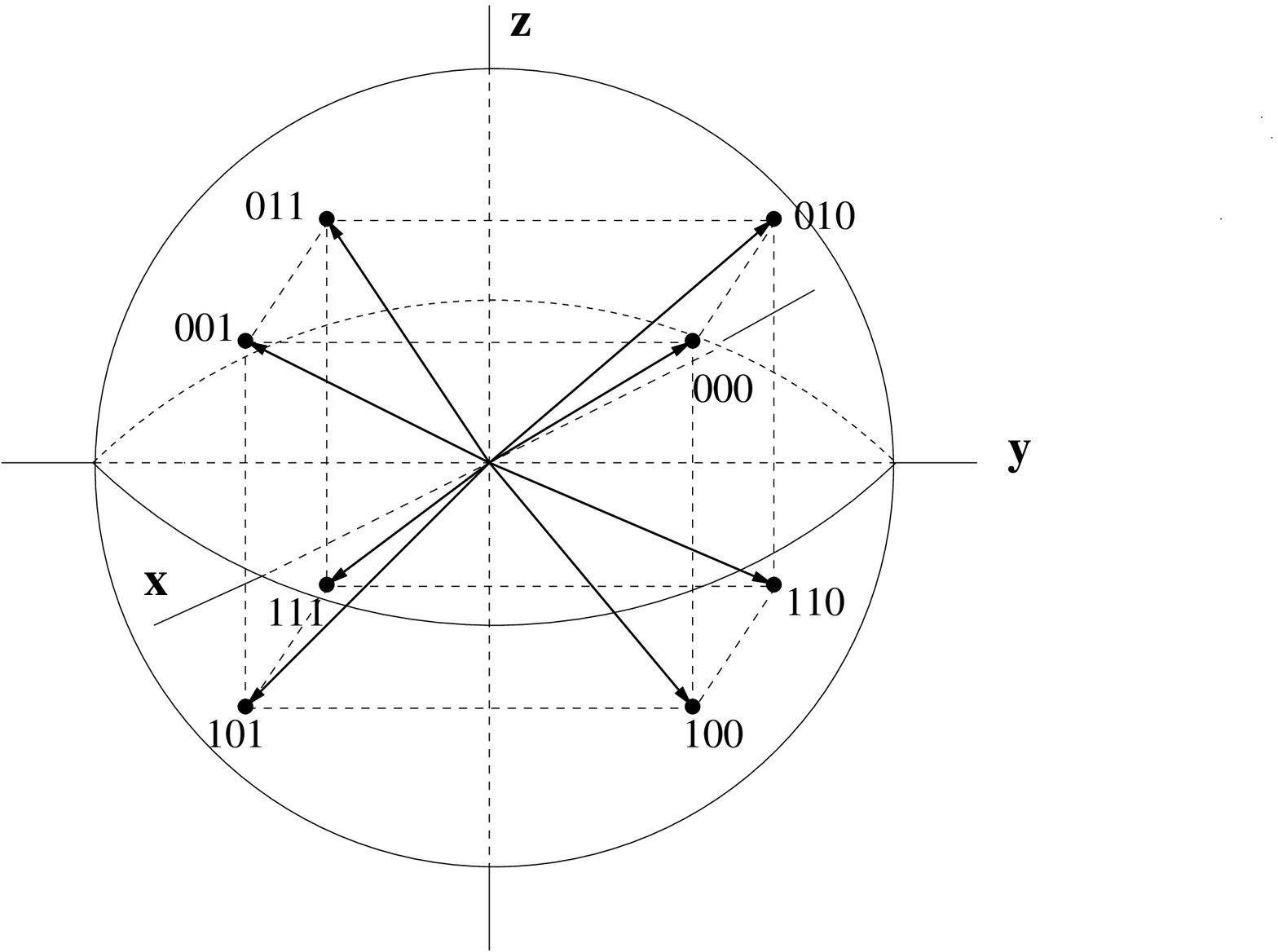}
\caption{The $(3,1,0.79)$-QRA coding in Bloch vector representation}\label{threetoone} 
\end{center}

\end{figure}

{\bf Example 1.} The $(2,1,0.85)$-QRA coding \cite{ANTV99} maps 
$x_1x_2\in \{0,1\}^2$ to $\pmb{\rho}_{x_1x_2} = 
\ket{\varphi(x_1x_2)}\bra{\varphi(x_1x_2)}$ where 
\[
\begin{array}{ll}
|\varphi(00)\rangle=\cos(\pi/8)|0\rangle+\sin(\pi/8)|1\rangle, 
&
|\varphi(10)\rangle=\cos(3\pi/8)|0\rangle+\sin(3\pi/8)|1\rangle,\\ 
|\varphi(11)\rangle=\cos(5\pi/8)|0\rangle+\sin(5\pi/8)|1\rangle, 
&
|\varphi(01)\rangle=\cos(7\pi/8)|0\rangle+\sin(7\pi/8)|1\rangle.
\end{array}
\] 
For decoding, we use the measurements by the following POVMs (projective measurements, in 
fact): $E^{1} = \{\ket{0}\bra{0},\ket{1}\bra{1}\}$,  and $E^{2} = 
\{\ket{+}\bra{+},\ket{-}\bra{-}\}$, where 
$|+\rangle=\frac{1}{\sqrt{2}}(|0\rangle+|1\rangle)$ and 
$|-\rangle=\frac{1}{\sqrt{2}}(|0\rangle-|1\rangle)$. See 
Fig.\ref{twotoone}. To decode the second bit, for example, 
we measure the encoding state in the basis $\{\ket{+},\ket{-}\}$. The angle between 
$\ket{\varphi(00)}$ and $\ket{+}$ (and also between 
$\ket{\varphi{(10)}}$ and $\ket{+}$) is $\pi/8$ and hence 
the success probability of decoding the value $0$ is $\cos^2{(\pi/8)} > 0.85$.  

To explain $(3,1,0.79)$-QRA coding, the Bloch sphere is convenient, 
which is based on the following two facts (see, e.g., \cite{NC00}): ($i$) 
let $\pmb{\rho}$ be any one-qubit quantum state, $\vec{r} = (r_x,r_y,r_z)$ be 
a (real) vector such that $|\vec{r}| \le 1$, and $X,Y,Z$ be $2\times 2$ Pauli matrices such that 
\begin{equation*}
X = 
\left(
\begin{array}{cc}
0 & 1\\
1 & 0
\end{array}
\right),\ \  
Y = 
\left(
\begin{array}{cc}
0 & -\imath\\
\imath & 0
\end{array}
\right),\ \  
Z = 
\left(
\begin{array}{cc}
1 & 0\\
0 & -1
\end{array}
\right).
\end{equation*}
Then $\pmb{\rho} = \frac{1}{2}\left(I + r_x X + r_y Y + r_z Z\right)$ defines 
a one-to-one mapping between $\pmb{\rho}$ and $\vec{r}$. The vector $\vec{r}$ is called 
the {\em Bloch vector} of $\pmb{\rho}$. It is well-known that $\pmb{\rho}$ is pure 
iff $|\vec{r}| = 1$. ($ii$) Let $\vec{s}$ be the Bloch vector of a pure 
state $\ket{\psi}\bra{\psi}$. Then $\bra{\psi} \pmb{\rho} \ket{\psi} = 
\frac{1}{2}\left(1 + \vec{r}\cdot \vec{s}\right)$. 
Namely, the probability calculation for a POVM can be done by using Bloch vectors. 

{\bf Example~2.} The $(3,1,0.79)$-QRA coding (attributed to Chuang in 
\cite{ANTV99}) maps $x_1x_2x_3\in\{0,1\}^3$ to $\pmb{\rho}_{x_1x_2x_3} = \ket{\varphi(x_1x_2x_3)}\bra{\varphi(x_1x_2x_3)}$, 
where 
\[
\begin{array}{ll}
|\varphi(000)\rangle= \cos\tilde{\theta}|0\rangle+e^{\pi\imath/4}\sin\tilde{\theta}|1\rangle, & 
|\varphi(001)\rangle= \cos\tilde{\theta}|0\rangle+e^{-\pi\imath/4}\sin\tilde{\theta}|1\rangle,\\ 
|\varphi(010)\rangle= \cos\tilde{\theta}|0\rangle+e^{3\pi\imath/4}\sin\tilde{\theta}|1\rangle, & 
|\varphi(011)\rangle= \cos\tilde{\theta}|0\rangle+e^{-3\pi\imath/4}\sin\tilde{\theta}|1\rangle,\\ 
|\varphi(100)\rangle= \sin\tilde{\theta}|0\rangle+e^{\pi\imath/4}\cos\tilde{\theta}|1\rangle, & 
|\varphi(101)\rangle= \sin\tilde{\theta}|0\rangle+e^{-\pi\imath/4}\cos\tilde{\theta}|1\rangle,\\ 
|\varphi(110)\rangle= \sin\tilde{\theta}|0\rangle+e^{3\pi\imath/4}\cos\tilde{\theta}|1\rangle, & 
|\varphi(111)\rangle= \sin\tilde{\theta}|0\rangle+e^{-3\pi\imath/4}\cos\tilde{\theta}|1\rangle, 
\end{array}
\]
such that $\tilde{\theta}$ satisfies $\cos^2\tilde{\theta}=1/2+\sqrt{3}/6 >  0.79$. 

As shown in Fig.\ref{threetoone}, Bloch vectors for those eight states 
are $(\pm 1/\sqrt{3},\pm 1/\sqrt{3}, \pm 1/\sqrt{3})$. For decoding, we 
use projective measurements in the bases  $\{|0\rangle,|1\rangle\}$, 
$\{|+\rangle,|-\rangle\}$ and $\{|+'\rangle,|-'\rangle\}$ for recovering 
the first, second and third bits, respectively, whose Bloch vectors are $\pm(0,0,1)$ (z-axis), $\pm(1,0,0)$ 
(x-axis), and $\pm(0,1,0)$ (y-axis), respectively. Here, $|+'\rangle=\frac{1}{\sqrt{2}}(|0\rangle+\imath|1\rangle)$ 
and $|-'\rangle=\frac{1}{\sqrt{2}}(|0\rangle-\imath|1\rangle)$. Thus, for 
example, if we measure $\ket{\varphi(001)}$ by $\{\ket{+},\ket{-}\}$, 
then the probability of getting the correct value $0$ for the second bit 
is $1/2 + \sqrt{3}/6 > 0.79$.  

Note that the success probability of the $(3,1,0.79)$-QRA coding is 
worse than that of the $(2,1,0.85)$-QRA coding, but is still quite high. 
Thus, it might be natural to conjecture that we still have room to 
encode four bits into one qubit. In fact, \cite{ANTV02} gives a statement 
of positive flavour. Before disproving this conjecture, let us look at 
the third example, which might seem to work as a $(4,1,>1/2)$-QRA coding. 

{\bf Example~3.} For encoding $x_1x_2x_3x_4 \in \{0,1\}^4$, we select 
$\ket{\varphi{(0x_1x_2)}}$ and $\ket{\varphi{(1x_3x_4)}}$ uniformly at 
random, where $\ket{\varphi(z_1z_2z_3)}$ is the same state as the one 
used in the $(3,1,0.79)$-QRA coding. For decoding, we first apply the 
universal cloning \cite{BH96} to the qubit ($\ket{\varphi{(0x_1x_2)}}$ 
or $\ket{\varphi{(1x_3x_4)}}$) and let $\pmb{\rho}_1$ and $\pmb{\rho}_2$ be the 
first and the second clones, respectively. If we want to get $x_1$ 
($x_2$, resp.), we apply the decoding process of $(3,1,0.79)$-QRA coding 
to recover the first bit of $\pmb{\rho}_1$. If the result is $0$, then by 
assuming that the transmitted qubit was $\ket{\varphi{(0x_1x_2)}}$, we 
recover the second (third, resp.) bit of $\pmb{\rho}_2$ again by 
$(3,1,0.79)$-QRA decoding process. Otherwise, i.e., if the result is $1$, 
then by assuming that the transmitted qubit was  
$\ket{\varphi{(1x_3x_4)}}$, we output the random bit ($0$ or $1$ with 
equal probability). Decoding $x_3$ or $x_4$ is similar and omitted. 

First of all, one should see that the above protocol completely follows 
the definition of the QRA coding: The encoding process maps $x_1x_2x_3x_4$ to a mixed state. 
The decoding process is a bit complicated, but it is well-known (e.g., 
\cite{NC00}) that such a physically realizable procedure can be 
expressed by a single POVM. Suppose that the receiver wants to get $x_1$ 
or $x_2$. Then note that $\ket{\varphi{(0x_1x_2)}}$ is sent with 
probability $1/2$ and if that is the case, the receiver can get a 
correct result with probability $p_0$ which is strictly greater than 
$1/2$. Otherwise, i.e., if $\ket{\varphi{(1x_3x_4)}}$ is sent, then the 
outcome is completely random. Thus the total success probability is 
$(p_0+1/2)/2 > 1/2$. Why is this argument wrong?  

\section{Main Result}
\begin{theorem}\label{thm1}
There exists no $(4,1,p)$-QRA coding with $p>1/2$. 
\end{theorem}

First let us return to Fig.~\ref{threetoone} to see how 
$(3,1,0.79)$-QRA coding works. Recall that the measurements for 
recovering $x_1$, $x_2$ and $x_3$ are all projective measurements. Now 
one should observe that each measurement corresponds to a plane in the 
Bloch sphere which acts as a ``boundary'' for the encoding states. For example, the 
measurement in the basis $\{\ket{0},\ket{1}\}$ corresponds to the 
$xy$-plane (States $\ket{0}$ and $\ket{1}$ correspond to $+z$ and $-z$ axes, respectively, on the sphere, 
which means that the measurement determines whether the encoding state lies above or under the $xy$-plane). 
Thus, the three planes corresponding to projective measurements of 
$(3,1,0.79)$-QRA coding divide the Bloch sphere into eight disjoint 
regions, each of which includes exactly one encoding state. 

Now suppose that there is a $(4,1,p)$-QRA coding whose decoding process 
is four {\it projective} measurements. Then, by a simple extension of 
the above argument, each measurement corresponds to a plane 
and the four planes divide the sphere into, say, $m$ regions. On the other hand, by 
definition we have $16$ encoding states and hence $m \ge 16$. 
(Otherwise, some two states fall into the same region, meaning the same 
outcome for those states, a contradiction.) However, it is well-known 
that a three-dimensional ball cannot be divided into $16$ (or more) 
regions by four planes. Thus, we are done if the decoding process is 
restricted to projective measurements. Due to its potential image of POVM, 
it seems unlikely that there exists a simple extension of this argument 
to the case of POVMs. A little surprisingly, there does. 

\begin{lemma}\label{lemma1} If there exists $(4,1,p)$-QRA coding with $p > 1/2$, then 
the three-dimensional ball can be divided into $16$ distinct 
regions by $4$ planes. 
\end{lemma}

\begin{proof} Suppose that $(4,1,p)$-QRA coding with $p > 1/2$ exists. 
Then by definition there are 16 encoding states $\{\pmb{\rho}_w \}_{w\in\{0,1\}^4}$ 
and 4 POVMs $\{E_0^i,E_1^i\}_{i\in\{1,2,3,4\}}$ such that 
$\mathrm{Tr}(E_0^i\pmb{\rho}_{w}) \ge p$ if $w_i = 0$ and 
$\mathrm{Tr}(E_0^i\pmb{\rho}_{w}) \le 1 - p$ if $w_i = 1$. 
As shown in the previous section, $E_0^i$ and $E_1^i$ can be written as:
\begin{eqnarray*}
E_0^i &=& \alpha_1^i \ket{u_i} \bra{u_i} + \alpha_2^i \ket{u_i^\bot}\bra{u_i^\bot} \label{defprom1}\\
E_1^i &=& (1 - \alpha_1^i) \ket{u_i} \bra{u_i} + (1 - \alpha_2^i) 
\ket{u_i^\bot}\bra{u_i^\bot} \label{defprom2}, 
\end{eqnarray*}
for $0 \le \alpha_2^i \le \alpha_1^i \le 1$ and orthogonal states 
$\ket{u_i}$ and $\ket{u_i^\bot}$. Thus, for all $i$, 
$\mathrm{Tr}(E_0^i\pmb{\rho}_w)$ can be written as: 
\begin{equation}\label{eqpovm}
\forall i\left[\alpha_1^i \bra{u_i}\pmb{\rho}_w\ket{u_i} + \alpha_2^i 
\bra{u_i^\bot}\pmb{\rho}_w\ket{u_i^\bot} \left\{
\begin{array}{ll}
> 1/2 & \mathrm{if}\ w_i = 0,\\
< 1/2 & \mathrm{if}\ w_i = 1
\end{array}
\right.
\right].
\end{equation}
Denoting the Bloch vectors of $\pmb{\rho}_w$ and 
$\ket{u_i}\bra{u_i}$ as $\vec{r}_w$ and $\vec{u}_i$, respectively, (\ref{eqpovm}) is rewritten as 
\begin{equation}\label{eqpovmbl}
\forall i\ \left[ \frac{\alpha_1^i + \alpha_2^i}{2} +  
\frac{\alpha_1^i-\alpha_2^i}{2}\cdot\vec{r}_w\cdot \vec{u}_i \left\{
\begin{array}{ll}
> {1}/{2}& \mathrm{if}\ w_i=0,\\
< {1}/{2}& \mathrm{if}\ w_i=1
\end{array}
\right.\right],
\end{equation}
by Fact~($ii$) on the Bloch sphere. (Note that the Bloch vector for 
$\ket{u_i^\bot}\bra{u_i^\bot}$ is $-\vec{u_i}$.) 
If we let $c_i = 1-(\alpha_1^i+\alpha_2^i)$ and $\vec{s_i}=(\alpha_1^i - \alpha_2^i)\cdot \vec{u_i}$, 
(\ref{eqpovmbl}) becomes the following simple linear inequalities for the fixed $\vec{s_i}$s. 
\begin{equation}\label{eqpovmbl2}
\forall i\ \left[ \vec{r}_w\cdot \vec{s}_i \left\{
\begin{array}{ll}
> c_i& \mathrm{if}\ w_i=0,\\
< c_i& \mathrm{if}\ w_i=1
\end{array}
\right.\right].
\end{equation}

Now, let $B$ be the set of all Bloch vectors. 
Let also $D_{s_i}^{(0)}$ and  $D_{s_i}^{(1)}$ be the subsets of $\mathbb{R}^3$ defined  
 by $D_{s_i}^{(0)}=\{\vec{r}\in B\mid \vec{r}\cdot \vec{s}_i >c_i\}$
and $D_{s_i}^{(1)}=\{\vec{r}\in B\mid \vec{r}\cdot \vec{s}_i <c_i\}$, respectively. 
By (\ref{eqpovmbl2}), all 16 subsets $D_w=D_{s_1}^{(w_1)}\cap D_{s_2}^{(w_2)}\cap D_{s_3}^{(w_3)}\cap D_{s_4}^{(w_4)}$ 
must not be empty. These subsets are the 16 non-empty regions of the ball 
divided by the 4 planes $\{\vec{r}\mid \vec{r}\cdot \vec{s}_i =c_i\}$.
\end{proof}
 
Lemma~\ref{lemma1} contradicts the following well-known geometric fact, which 
completes the proof of Theorem~\ref{thm1}. 

\begin{lemma}\label{lemma2} 
A ball cannot be divided into 16 non-empty regions by 4 planes. 
\end{lemma}

By using the notion of Bloch vectors of $n$-qubit states, we have the following generalization. 
 
\begin{theorem}\label{thm2}
There is no $(2^{2m},m,p)$-QRA coding with $p>1/2$. 
\end{theorem}

The proof of Theorem \ref{thm2} proceeds similarly to Theorem \ref{thm1} 
except for the generalization of Bloch vectors and Lemma \ref{lemma2}. 
For completeness we repeat such a similar argument. 
First, we review the Bloch vectors of $N$-level quantum states \cite{AL87,JS01,Kim03,MW95}.   
Let $\pmb{\lambda}_1,\ldots,\pmb{\lambda}_{N^2-1}$ be orthogonal generators of $SU(N)$ 
satisfying: (i) $\pmb{\lambda}_i^\dagger=\pmb{\lambda}_i$; (ii) $\mathrm{Tr}(\pmb{\lambda}_i)=0$; 
(iii) $\mathrm{Tr}(\pmb{\lambda}_i\pmb{\lambda}_j)=2$ if $i=j$ and $0$ if $i\neq j$. Then, any 
$N$-level quantum state $\pmb{\rho}$ can be represented as an $(N^2-1)$-dimensional real vector 
$\vec{r}=(r_1,\ldots,r_{N^2-1})$, called the Bloch vector of $\pmb{\rho}$, such that $\pmb{\rho}=\frac{1}{N}I_N+\frac{1}{2}\sum_{i=1}^{N^2-1} r_i\pmb{\lambda}_i$, 
where $I_N$ is the $N$-dimensional identity matrix. 
Note that, by the properties of $\pmb{\lambda}_i$s, for any two $N$-level quantum states $\pmb{\rho}$ 
and $\pmb{\sigma}$ with their Bloch vectors $\vec{r}$ and $\vec{s}$, 
\begin{equation}\label{Bloch}
\mathrm{Tr}(\pmb{\rho}\pmb{\sigma})=\frac{1}{N}+\frac{1}{2}\cdot\vec{r}\cdot\vec{s}.
\end{equation}
Second, we give the following lemma.  

\begin{lemma}\label{lemma3} If there exists a $(2^{2m},m,p)$-QRA coding with $p>1/2$, 
then $\mathbb{R}^{2^{2m}-1}$ can be divided into $2^{2^{2m}}$ distinct regions 
by $2^{2m}$ hyperplanes. 
\end{lemma}

\begin{proof} Suppose that $(2^{2m},m,p)$-QRA coding with $p>1/2$ exists. 
Then, by definition there are $2^{2^{2m}}$ encoding states $\{\pmb{\rho}_w\}_{w\in\{0,1\}^{2^{2m}}}$ 
and $2^{2m}$ POVMs $\{E_0^i,E_1^i\}_{i\in\{1,2,\ldots,2^{2m} \}}$ such 
that $\mathrm{Tr}(E_0^i\pmb{\rho}_w) > 1/2$ if $w_i=0$ 
 and $\mathrm{Tr}(E_0^i\pmb{\rho}_w) < 1/2$ if $w_i=1$. 
Since $E_0^i$ and $E_1^i$ are $2^m$-dimensional nonnegative Hermitian, they can be written as:
\begin{eqnarray*}
E_0^i &=& \sum_{j=1}^{2^{m}} \alpha_j^i \ket{u_j^i} \bra{u_j^i} \\
E_1^i &=& \sum_{j=1}^{2^{m}} (1 - \alpha_j^i) \ket{u_j^i} \bra{u_j^i}, 
\end{eqnarray*}
such that $\{\ket{u_j^i}\}_{j=1}^{2^m}$ is an orthonormal basis. 
Thus, for all $i\in\{1,\ldots,2^{2m}\}$, the following must be satisfied: 
\begin{equation}\label{41-1}
\forall i\ \left[ 
\sum_{j=1}^{2^m} \alpha_j^i \langle u_j^i|\pmb{\rho}_w| u_j^i\rangle
\left\{
\begin{array}{ll}
>1/2 & \mathrm{if}\ w_i=0,\\
<1/2 & \mathrm{if}\ w_i=1
\end{array}
\right.
\right].
\end{equation}
Denoting the Bloch vectors of $\pmb{\rho}_w$ and $\ket{u_j^i}\bra{u_j^i}$ as $\vec{r}_w$ and $\vec{u}_j^i$ 
(which are $(2^{2m}-1)$-dimensional real vectors), respectively, (\ref{41-1}) is rewritten as
\begin{equation}\label{41-2}
\forall i\ \left[ 
\sum_{j=1}^{2^m} ( 
\frac{\alpha_j^i}{2^m}  + \frac{\alpha_j^i}{2} \vec{r}_w\cdot \vec{u}_j^i )
\left\{
\begin{array}{ll}
> 1/2& \mathrm{if}\ w_i=0,\\
< 1/2& \mathrm{if}\ w_i=1
\end{array}
\right.\right]
\end{equation}
by (\ref{Bloch}). (Notice that an $m$-qubit state can be identified with a $2^m$-level quantum state.) 
If we let $c_i=1/2-\sum_{j=1}^{2^m} \frac{\alpha_j^i}{2^m}$ 
     and $\vec{s_i}=\sum_{j=1}^{2^m} \frac{\alpha_j^i}{2}\vec{u}_j^i$, (\ref{41-2}) is simplified as follows:
\begin{equation}\label{41-3}
\forall i\ \left[ 
\vec{r}_w\cdot\vec{s_i}
\left\{
\begin{array}{ll}
> c_i& \mathrm{if}\ w_i=0,\\
< c_i& \mathrm{if}\ w_i=1
\end{array}
\right.\right].
\end{equation}

Let $B$ be the set of all Bloch vectors for $m$-qubit states. 
Note that $B\subseteq \mathbb{R}^{2^{2m}-1}$. 
Let also $D_{s_i}^{(0)}$ and $D_{s_i}^{(1)}$ be the subsets of $\mathbb{R}^{2^{2m}-1}$ 
defined by $D_{s_i}^{(0)}=\{\vec{r}\in B\mid \vec{r}\cdot \vec{s}_i > c_i\}$
and        $D_{s_i}^{(1)}=\{\vec{r}\in B\mid \vec{r}\cdot \vec{s}_i < c_i\}$, respectively. 
By (\ref{41-3}), all $2^{2^{2m}}$ subsets $D_w=\bigcap_{i=1}^{2^{2m}} 
D_{s_i}^{(w_i)}$ must not be empty. These subsets are included into non-empty regions 
of $\mathbb{R}^{2^{2m}-1}$ divided by the $2^{2m}$ hyperplanes $\{\vec{r} \mid \vec{r}\cdot \vec{s}_i = c_i\}$. 
\end{proof}

Now, the following geometric fact (see, e.g., \cite{Ede87}) completes the proof of Theorem \ref{thm2}. 

\begin{lemma}\label{lemma4} 
$\mathbb{R}^{2^{2m}-1}$ cannot be divided into $2^{2^{2m}}$ 
non-empty regions by $2^{2m}$ hyperplanes. 
\end{lemma}

\section{Applications to Network Coding}
\begin{figure}[bt]
\begin{center}
\includegraphics*[width=4.0cm]{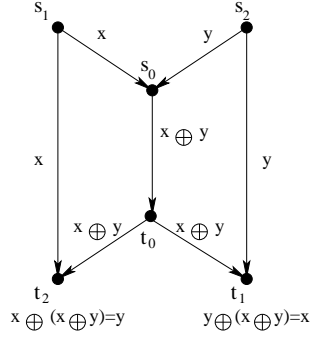}
\caption{Butterfly network}\label{classicalnc} 
\end{center}
\end{figure}

Network coding, introduced in \cite{ACLY00}, is nicely explained by using the so-called Butterfly network 
as shown in Fig.~\ref{classicalnc}. The capacity of each directed link is all
one and there are two source-sink pairs $s_1$ to $t_1$ and $s_2$ to $t_2$. 
Notice that both paths have to use the single link from $s_0$ to $t_0$ 
and hence the total amount of flow in both paths is bounded by one, say, $1/2$ for each. 
Interestingly, this max-flow min-cut theorem no longer applies for ``digital information flow.''  
As shown in the figure, we can transmit two bits, $x$ and $y$, on the two paths simultaneously.

The paper \cite{HINRY06} extends this network coding for quantum channels and 
quantum information. Their results include; 
($i$) One can send any quantum state $|\psi_1\rangle$ 
from $s_1$ to $t_1$ and $|\psi_2\rangle$ from $s_2$ to $t_2$ simultaneously 
with a fidelity strictly greater than $1/2$. 
($ii$) If one of $|\psi_1\rangle$ and $|\psi_2\rangle$ is classical, 
then the fidelity can be improved to $2/3$. 
($iii$) Similar improvement is also possible if $|\psi_1\rangle$ and $|\psi_2\rangle$ 
are restricted to only a finite number of (previously known) states. 
This allows us to design a protocol which can send three classical bits 
from $s_1$ to $t_1$ (similarly from $s_2$ to $t_2$) but only one of them should be recovered.

By our result in this paper, we can prove a kind of optimality of the 
result ($iii$). Firstly, we cannot extend the above three bits to four 
bits. The reason is easy: if we could then we would get a 
$(4,1,>1/2)$-QRA coding for the $s_1$--$t_1$ path by fixing the state at 
$s_2$ to say $\ket{0}$. Secondly, we can prove that the two side links 
($s_1$ to $t_2$ and $s_2$ to $t_1$) which are unusable in the 
conventional multicommodity flow are in fact useful; if we remove them, 
then the network can be viewed as a $(4,1,p)$-QRA coding system, which 
cannot achieve $p > 1/2$.

\section{Concluding Remarks}
An interesting open question is the possibility of $(n,2,>1/2)$-QRA 
coding. $(6,2,0.79)$-QRA coding is obvious since we can use two 
$(3,1,0.79)$-QRA codings independently. For $n=7$, there is the 
following simple construction.

{\bf Example~4.} The $(7,2,0.54)$-QRA coding consists of encoding states 
and measurements as follows. For each seven bits 
$x=x_1x_2x_3x_4x_5x_6x_7$, the encoding state $\pmb{\rho}(x)$ is  
\[
\alpha \ket{\varphi(x_1x_2x_3)}\bra{\varphi(x_1x_2x_3)}\otimes\ket{\varphi(x_4x
_5x_6)}\bra{\varphi(x_4x_5x_6)} + (1-\alpha)\ket{\xi(x_7)}\bra{\xi(x_7)} 
\]
with $\alpha=\frac{6}{7+\sqrt{3}}$, 
where $\ket{\xi(0)}=\frac{1}{\sqrt{2}}(\ket{00}+\ket{11})$ 
  and $\ket{\xi(1)}=\frac{1}{\sqrt{2}}(\ket{01}+\ket{10})$. To obtain 
any one of $x_1,x_2$ and $x_3$ (resp.\ $x_4,x_5$ and $x_6$)  
use the measurement of the $(3,1,0.79)$-QRA coding on the first qubit 
(resp.\ second qubit) of $\pmb{\rho}(x)$. To obtain $x_7$, use the projective measurement 
$E^7 = \{E_0^7,E_1^7\}$ on $\pmb{\rho}(x)$, where $E_0^7=\ket{00}\bra{00}+\ket{11}\bra{11}$ 
and $E_1^7=\ket{01}\bra{01}+\ket{10}\bra{10}$. Details are omitted.

\end{document}